\documentclass[pre,showpacs]{revtex4}
\usepackage{graphicx,bm}
\begin{document}

\title{Scar Intensity Statistics in the Position Representation}
\author{K. Damborsky and L. Kaplan}
\affiliation{Department of Physics, Tulane University, New Orleans,
Louisiana 70118, USA}
\date{October 6, 2005}

\begin{abstract}
We obtain general predictions for the distribution of wave function intensities in
position space on the periodic orbits of chaotic ballistic systems.  The
expressions depend on effective
system size $N$, instability exponent $\lambda$ of the periodic orbit,
and proximity to a focal point of the orbit. Limiting expressions are obtained that include
the asymptotic probability distribution of rare high-intensity events and a perturbative
formula valid in the limit of weak scarring.  For finite system sizes, a single scaling
variable $\lambda N$ describes deviations from the semiclassical $N \to \infty$ limit.
\end{abstract}


\maketitle

\section{Introduction}
\label{secintro}

Improving our understanding of wave function structure and transport
in quantum systems with a non-integrable classical limit, and of
their relationship with the dynamics of the corresponding classical
system, has been a major focus of quantum chaos research.  An early
and visually dramatic finding along these lines was the presence of
anomalously large fluctuations in wave function intensity near the
short unstable periodic orbits of the corresponding classical
dynamics.  In 1983, McDonald~\cite{mcdonald} observed individual
wave functions (later called ``scarred states") with a large excess
of intensity near a particular periodic orbit, as compared with the
background Gaussian random fluctuations; the completeness condition
then requires other wave functions (the ``anti-scarred states") to
have deficient intensity near the same orbit. Since that time, scars
have been recognized as the one of the few visually distinctive
non-random features of quantum chaotic wave functions, as well as
possibly the leading correction to the random wave conjecture made
earlier by Berry and Voros~\cite{randwave}, which states that
quantum wave function statistics should converge to the behavior of
random superpositions of plane waves in the semiclassical limit of
short wavelength.

The original theory of scarring was developed by
Heller~\cite{hellerscar}, who placed a lower bound on scar
fluctuations, as measured by overlaps of wave functions with
Gaussian wave packets centered on the periodic orbits.  Subsequent
theoretical successes included work by Bogomolny~\cite{bogoscar},
who first computed an expression for scar intensity in the position
representation (where scars were originally observed), and by
Berry~\cite{berryscar}, who developed a parallel analysis for the
Wigner functions in phase space and predicted the structure of
fringe patterns decorating each orbit.  In these pioneering
investigations, the focus was mostly on average intensities for an
energy window of nearby eigenstates, rather than on properties of
individual wave functions.  Later, Agam and
Fishman~\cite{agamfishman} developed a semiclassical criterion for
predicting scarring or anti-scarring in individual wave functions.
In the semiclassical limit, the phase space region affected by the
scar of any given periodic orbit tends to zero while the intensity
enhancement in that region for a typical wave function remains
finite; the scar effect is therefore fully consistent with the
quantum ergodicity theorems of Shnirelman, Zelditch, and Colin de
Verdiere~\cite{shnir}.

Scars have been predicted and observed both experimentally and
numerically in a wide variety of quantum systems with a chaotic
classical limit (as well as in classical wave systems with a chaotic
ray limit).  Examples include applications as diverse as microwave
cavities~\cite{sridharscar}, hydrogen atoms in a magnetic
field~\cite{wintgen}, electrons in a resonant tunneling diode with
magnetic field~\cite{wilkinson}, Faraday surface
waves~\cite{faraday}, vibrating soap films~\cite{soap}, molecular
vibrations~\cite{molec}, and acoustic radiation~\cite{acoustic}.

In quantum or classical wave applications where dissipation effects
are small enough for individual eigenstates to be resolved, it is of
primary interest to study the scar effect for individual wave
functions, rather than for averages over an energy or frequency
window.  In fact, the early empirical evidence for scarring as well
as much of the empirical evidence since has been provided in the
form of images of individual eigenstates.  Such evidence, however,
needs to be interpreted with care in light of the fact that random
waves also exhibit regions of high intensity that to the eye look
like scars but are in reality perfectly consistent with Gaussian
random amplitude fluctuations~\cite{scarlet}.  To understand the
effect of scarring on physical processes in specific systems, and
even to verify the presence of this phenomenon in the first place,
it is necessary to have robust quantitative predictions concerning
the statistical properties of individual wave functions, which can
be compared with data as well as with the Gaussian random model.

Antonsen {\it et al.}~\cite{antonsen} noted that scar statistics of
individual wave functions could be understood by combining known
short-time dynamics associated with a given periodic orbit and
random long-time recurrences whose statistical properties mirror
those of random matrix theory (RMT).  Similar ideas were used by
Kaplan and Heller to obtain predictions for mean squared intensity
on an orbit~\cite{kaplanheller}, and also for the full distribution
of wave function intensities on or away from the
orbit~\cite{kaplaninten}. These predictions were obtained in a
Gaussian (Husimi) phase space basis, where classical--quantum
correspondence is most clearly manifest.  On the other hand, wave
function structure in position space is often of greater physical
interest.  Indeed, the original empirical evidence for
scarring~\cite{mcdonald} and much subsequent numerical and
experimental investigation of scarring has focused on position-space
properties, in contrast with the above-mentioned theoretical work.
Wave function intensities in position space are directly accessible,
for example, in microwave cavity experiments, and in the case of a
diffusive billiard, comparisons have been made with predictions
based on nonlinear sigma models~\cite{sridhar}.  Such explicit
comparisons between observed deviations from RMT and theoretical
predictions are lacking in the ballistic case.  Furthermore, a
quantitative understanding of position-space wave function
statistics is essential for analyzing experiments on nanostructures
such as quantum dots.  For example, these statistics determine the
effects of electron--electron interactions on conductance peak
spacings in the Coulomb blockade regime~\cite{cbthy}.

This paper is organized as follows.  The inverse participation
ratio, or mean squared wave function intensity, on a periodic orbit
is first obtained in Section~\ref{secipr}, followed by an analysis
of the full wave function intensity distribution in
Section~\ref{secdistr}.  In both cases, accurate and robust
expressions for these statistical properties are obtained that
depend only on the system size, instability exponent of the periodic
orbit and proximity to a focal point.  Several limiting expressions
are obtained, including a result for the probability of rare
high-intensity events, and a perturbative formula valid for moderate
intensities when deviations from RMT are small.  Finally, in
Section~\ref{secnumer}, we compare these results with data obtained
in numerical simulations, and briefly investigate the behavior of
wave function statistics beyond the semiclassical regime.

\section{Inverse Participation Ratio}
\label{secipr}

We begin by considering an unstable periodic orbit of period $T$ in $d$
dimensions, passing through periodic point $(q_0,p_0)$.  The
Van-Vleck--Gutzwiller semiclassical propagator~\cite{gutz} includes a
contribution from this periodic orbit to the return amplitude at time $t=nT$
for position state $q_0$:
\begin{equation}
\label{gutz}
G^{\rm SC}(q_0,q_0,t)={1 \over (2\pi i \hbar)^{d/2}}
\left |{\rm det} \;{\partial p(q_0,q_t) \over \partial q_t}\right|_{q_t=q_0}^{1/2}
e^{in(S/\hbar-\mu\pi/2)} + \ldots \,.
\end{equation}
Here $S$ is the classical action associated with one traversal of
the periodic orbit, and $\mu$ is the Maslov index, absorbing changes
in the sign of the amplitude.  The amplitude $|{\rm det} \,\partial
p(q_0,q_t)/ \partial q_t|_{q_t=q_0}^{1/2}$, where $p$ is the initial
momentum needed to travel from fixed $q_0$ to $q_t$ in time $t$, is
evaluated for the return trajectory $q_t=q_0$, and is the square
root of the corresponding classical focusing factor.  Of course,
$q_0$, $p$, and $q_t$ are all $d-$component vectors in the general
case.  Omitted terms in Eq.~(\ref{gutz}) are associated with other
classical paths of length $t$ beginning and ending at the same point
$q_0$; as we will see below the periodic path dominates the
statistics of individual wave functions when the instability
exponent of this periodic path is small.  For sufficiently short
times $t$ or sufficiently small $\hbar$, $G^{\rm SC}(q_0,q_0,t)$ is
a good approximation to the true quantum return amplitude
$G(q_0,q_0,t)$.

For simplicity of presentation, and for easy comparison with numerical results
in Section~\ref{secnumer}, we now restrict ourselves to the case of
discrete-time maps, which are commonly used as models for more general chaotic
dynamical systems~\cite{mapmodels}.  One-dimensional maps may be obtained by
stroboscopically viewing a one-dimensional system with time-dependent
Hamiltonian (such as the kicked rotator or standard map~\cite{standmap}), or
alternatively by taking a Poincar\'e surface of section for a two-dimensional
time-independent Hamiltonian (such as a hard-wall billiard or smooth
two-dimensional potential).  For a one-dimensional map on a compact phase space
of area $1$, the quantum vector space is spanned by a finite basis of
$N=1/2\pi\hbar$ independent position states $|q_i\rangle$.  Conventionally, we
adopt a wave function normalization where the average of the discrete wave
function intensities is set to unity: $\sum_{i=0}^{N-1} |\Psi(q_i)|^2=N$, and
the propagator $G(q_t,q,t)$ becomes a unitary $N$ by $N$ matrix.  With this
normalization,
\begin{equation}
\label{gutzmap}
G^{\rm SC}_{\rm map}(q_0,q_0,t)=\left\{ \begin{array}{ll}{1 \over (iN)^{1/2}}
\left |{\partial p(q_0,q_t) \over \partial q_t}\right|_{q_t=q_0}^{1/2}
e^{in(S/\hbar-\mu\pi/2)} + \ldots & n > 0 \\
1 & n =0 \end{array} \right. \,,
\end{equation}
where $n=t/T$ is an integer.  Furthermore, for a $d=1$ map, or equivalently for
a $d=2$ continuous-time dynamics, the monodromy matrix of the unstable periodic
orbit is simply a 2 by 2 matrix, and may be written as
\begin{equation}
\label{monodromy}
\left( \begin{array}{cc} \partial q_t(q,p)/\partial q & \partial q_t(q,p)/\partial
p \\ \partial p_t(q,p)/\partial q & \partial p_t(q,p)/\partial p
\end{array} \right )
= \left( \begin{array}{cc} a & b \\ c & d \end{array} \right )
\left ( \begin{array}{cc} e^{n\lambda } & 0 \\ 0 & e^{-n\lambda } \end{array}
\right ) \left( \begin{array}{cc} d & -b \\ -c & a \end{array} \right ) \,,
\end{equation}
where all partial derivatives are evaluated at the periodic point $(q,p)=(q_0,p_0)$,
area preservation implies
$ad-bc=1$, and $\lambda>0$ is the dimensionless positive instability exponent for one
iteration of the periodic orbit.  Then the inverse of the focusing factor in
Eq.~(\ref{gutzmap}) becomes
\begin{equation}
\label{focfact}
(|\partial p(q_0,q_t)/\partial q_t|_{q_t=q_0})^{-1} = |\partial q_t(q_0,p)/\partial
p|_{q_t=q_0}=2|ab\,\sinh(\lambda t/T)|\,.
\end{equation}

We are interested in the effect of the unstable periodic orbit on individual
wave function behavior at or near $q_0$, and specifically on the wave function
intensities $|\Psi(q_0)|^2$.  A general discussion of the intensity distribution
${\cal P}(|\Psi(q_0)|^2)$ is deferred until Section~\ref{secdistr}; here we
focus first on the mean squared intensity, also known as the inverse
participation ratio (IPR),
\begin{equation}
{\cal I}(q_0)= \overline{|\Psi(q_0)|^4}\,,
\end{equation}
where the average $\overline{\cdots}$ is performed over eigenstates of the
system, and the mean intensity $\overline{|\Psi(q_0)|^2}$ is normalized to
unity as above.  For a general Hamiltonian system, the average must be
restricted to a classically narrow energy window; in the case of a
discrete-time map, however, we are free to average over all values of the
(periodic) quasi-energy.  ${\cal I}(q_0)$ is the inverse fraction of wave
functions in the energy window that have significant intensity at $q_0$; its
possible values range from $1$ in the case of complete ergodicity (where every
wave function has equal intensity at $q_0$) to $N$ in the case of complete
localization (where the wave functions are delta functions in position space).

As has previously been discussed~\cite{iprst}, for a non-degenerate system the
IPR is proportional to the long-time return probability for initial state
$|q_0\rangle$, which for the chaotic case is in turn proportional to a sum of
return probabilities due to the periodic orbit,
\begin{eqnarray}
{\cal I}(q_0) &=& {\cal N} \int_{-\infty}^\infty dt \, |G_{\rm map}
(q_0,q_0,t)|^2 \nonumber \\
\label{iprgen}
&\approx & F \, {\sum_{n=-\infty}^\infty |G_{\rm map}^{\rm p.o.}(q_0,q_0,nT)|^2
\over |G^{\rm p.o.}_{\rm map}(q_0,q_0,0)|^2} \,.
\end{eqnarray}
Here $G^{\rm p.o.}_{\rm map}$ is the contribution to the propagator
from a single dominant periodic orbit, as given in Eq.~(\ref{gutz}),
and ${\cal N}$ is a normalization constant.  The multiplicative
factor $F$ encodes information about all recurrences not included in
$G^{\rm p.o.}_{\rm map}$; semiclassically these recurrences are
associated with closed classical trajectories other than the
original periodic orbit.  Treating these other contributions as
random and uncorrelated, which is physically justifiable for chaotic
dynamics at long times, leads to the random wave result $F=3$ (for
real wave functions, in the presence of time reversal symmetry) or
$F=2$ (for complex wave functions, in the absence of time reversal
symmetry).

Combining the results of Eqs.~(\ref{iprgen}), (\ref{gutzmap}), and
(\ref{focfact}), we obtain the prediction
\begin{equation}
{\cal I}_{\rm pred}(q_0) = F \left [ 1+ \sum_{n \ne 0} { 1\over 2 N|ab
\,\sinh(\lambda n)|} \right ]= F\left [ 1+ {1 \over  N|ab|}
\sum_{n=1}^\infty { 1 \over \sinh(\lambda n)} \right ] \,.
\label{scaripr}
\end{equation}
For instability exponent $\lambda \gg 1$, only one iteration of the orbit
contributes, and we obtain ${\cal I}_{\rm pred}(q_0)=F \left [1+{2 \over
N|ab|}e^{-\lambda}\right ]$, converging eventually to the random wave result
${\cal I}_{\rm RMT}(q_0)=F$.  In the more interesting limit $\lambda \ll 1$, we
obtain \begin{equation}
\sum_{n=1}^\infty { 1 \over \sinh(\lambda n)}  \approx
\sum_{n=1}^M { 1 \over \lambda n} +
\int_M^\infty {dn \over \sinh(\lambda n)}  \approx
{1 \over \lambda} \left ( \ln{1 \over \lambda} +\gamma + \ln 2 \right ) \,,
\end{equation}
where $1 \ll M \ll \lambda^{-1}$ and $\gamma \approx 0.577$ is Euler's
constant.  Then we find
\begin{equation}
{\cal I}_{\rm pred}(q_0) = F \left [ 1+{1 \over  \lambda|ab|N}  \left(
\ln{2 \over \lambda} +\gamma \right ) \right ]
\label{smlamipr}
\end{equation}
in the weakly unstable regime $\lambda \ll 1$.

From Eq.~(\ref{scaripr}) or Eq.~(\ref{smlamipr}), we see that scar strength, as
measured by the inverse participation ratio, depends on three parameters only:
the semiclassical parameter $N$ (or ratio of system size to wavelength), the
instability exponent $\lambda$ of the periodic orbit, and the phase space
orientation parameter $|ab|$, which is directly related to the angle between
the $p-$axis and the unstable manifold of the orbit at the point $(q_0,p_0)$.
In a two-dimensional Hamiltonian system, such as a billiard or smooth
potential, the stable manifold rotates and the $|ab|$ parameter changes as one
moves along the orbit, approaching zero at the focal points.  Thus, all
possible values of $|ab|$ are relevant for describing the wave function
behavior near a generic orbit, while $\lambda$ is a constant parameter for a
fixed orbit.  For small values of $|ab|$, which correspond to the strongest
scarring, $|ab|$ is directly proportional to the distance from the nearest
focal point on the orbit.  In a specific discrete-time map, $|ab|$ has a single
value for each orbit of period one, since each such orbit has only one periodic
point.  However, the same correspondence between the $|ab|$ parameter and
distance to a focal point applies also in the case of maps, if we consider a
{\it family} of maps arising from Poincar\'e sections intersecting a given
orbit at various distances from a focal point.

We need to understand the range of validity for Eq.~(\ref{scaripr})
or Eq.~(\ref{smlamipr}).  Classical--quantum correspondence
$G(q_0,q_0,t) \approx G^{\rm SC}(q_0,q_0,t)$ fails when $q_0$ is
within a wavelength of a focal point, as the classically singular
probability density is smoothed out by the uncertainty principle.
Thus, our semiclassical derivation fails for $|ab| < N^{-1}$, and
the correct scaling behavior for this regime must be obtained by
replacing $|ab|N$ with a number of order unity.  Thus, the inverse
participation ratio ${\cal I}(q_0)$ remains finite at
$O(\lambda^{-1})$ for $\lambda \ll 1$ as $|ab|N \to 0$.  We note
that the IPR scales as $\lambda^{-1}$ in the basis of phase-space
Gaussians optimally oriented with respect to the stable and unstable
manifolds of the periodic orbit~\cite{kaplanheller}; it is not
surprising to obtain the same scaling behavior in position space
near the focal points of the orbit, since a position state in this
case may be thought of simply as a limiting case of a family of
optimal Gaussians.

A second limitation on the validity of Eq.~(\ref{scaripr}) or
Eq.~(\ref{smlamipr}) is that the Lyapunov time $T/\lambda$ needed to escape
from the vicinity of the orbit must be much shorter than the Heisenberg time
$N$, at which quantum mechanical exploration ceases and the quantum dynamics
becomes quasi-periodic.  For $\lambda N< T$, the infinite sum in
Eq.~(\ref{scaripr}) must be cut off at $n \sim N/T$ to avoid counting classical
recurrences that occur after the Heisenberg time and have no quantum analogue.
This has the effect of replacing $\lambda N$ in Eq.~(\ref{smlamipr}) with a
constant of order $T$ when $\lambda N <T$, so that the inverse
participation ratio again remains finite: ${\cal I}(q_0) \sim |ab|^{-1}T^{-1}$ as
$\lambda \to 0$.  The behavior of the IPR for $\lambda N <T$ will be discussed
further in Section~\ref{secnumer}.  For $\lambda \to 0$ and $|ab| \to 0$
simultaneously, the IPR must obey an upper bound ${\cal I}(q_0) \le N$,
which is saturated only if the position state $|q_0\rangle$ is itself
an eigenstate of the dynamics.

In conclusion, for a given semiclassical parameter $N$, we require
$|ab|>N^{-1}$ and $\lambda >T N^{-1}$ for our semiclassical
expressions to be valid; smaller values of $\lambda$ or $|ab|$ do
not lead to parametrically stronger scarring.  At the same time, as
seen from Eq.~(\ref{smlamipr}), we need simultaneously $\lambda  \ll
1$ and $|ab| \ll 1$ to obtain a large enhancement of the IPR above
the RMT prediction.  We also note that in any specific dynamical
system, multiple trajectories (periodic or closed) contribute to
${\cal I}(q_0)$ at order $N^{-1}$, and a single orbit can only be
expected to dominate the wave function statistics at $q_0$ when
$|ab| \ll 1$ and $\lambda \ll 1$, i.e. when $q_0$ is close to a
focal point of a weakly unstable periodic orbit.

\section{Scar Intensity Distribution}
\label{secdistr}

Although the IPR provides a useful one-number measure of the degree
of wave function localization at $q_0$, it is only the lowest
non-trivial moment of the intensity distribution, and as such
provides limited information about the full distribution.  In
particular, ${\cal I}(q_0)>{\cal I}_{\rm RMT}=F$ implies an
increased intensity variance and clearly suggests longer tails of
the intensity distribution, i.e. an enhanced probability of finding
wave functions with intensity $I=|\Psi(q_0)|^2$ much larger or much
smaller than the average intensity.  However, to obtain quantitative
predictions about the probability of such rare events, we must go
beyond the IPR to examine the full intensity distribution ${\cal
P}(I)$.

As in Section~\ref{secipr}, we assume a clear separation of scales
exists between known short-time recurrences associated with a
particular short periodic orbit and new long-time nonlinear
recurrences which semiclassically may be associated with an
exponentially large number of homoclinic paths starting and ending
near $q_0$.  The statistical properties of these new recurrences at
long times may be assumed to be consistent with RMT, i.e.
independent of the fact that $q_0$ happens to lie on a short
periodic orbit~\cite{kaplanheller}. The full return amplitude at
long times then becomes a convolution of the known short-time
dynamics and the RMT-like random long-time
recurrences~\cite{kaplanheller}:
\begin{equation}
G_{\rm map}(q_0,q_0,t) \approx \sum_\tau G_{\rm map}^{\rm
p.o.}(q_0,q_0,\tau)G^{\rm rnd}(q_0,q_0,t-\tau) \,.
\end{equation}
Fourier transforming, we find that the local density of states at $q_0$ is
given by the product of a smooth envelope associated with the periodic orbit
and a random fluctuating part:
\begin{equation}
\sum_\alpha |\Psi_\alpha(q)|^2 \, \delta (E-E_\alpha)=
\tilde G_{\rm map}(q_0,q_0,E)= \tilde G_{\rm map}^{\rm
p.o.}(q_0,q_0,E) \sum_\alpha |r_\alpha|^2 \delta (E-E_\alpha) \,,
\end{equation}
where the $r_\alpha$ are distributed as (real or complex) independent Gaussian
random variables with mean $0$ and variance $1$, and $G_{\rm map}^{\rm
p.o.}(q_0,q_0,E)$ is the Fourier transform of Eq.~(\ref{gutzmap}).  The
individual wave function intensities are given by
\begin{equation}
I_\alpha=|\Psi_\alpha(q_0)|^2=\tilde G_{\rm map}^{\rm p.o.}(q_0,q_0,E_\alpha)
|r_\alpha|^2 =S(E_\alpha) |r_\alpha|^2 \,,
\label{infact}
\end{equation}
where we have given the name $S(E)$ to the smooth part of the local density of
states at $q_0$.  Explicitly, this smooth envelope coming from the short
periodic orbit takes the form
\begin{equation}
\label{se}
S(E)=1+ 2 {1 \over \sqrt {2N|ab|}} \sum_{n=1}^\infty  {1 \over
\sqrt{\sinh(\lambda n)}} \cos {\left({n(ET+S)\over\hbar}-{n\mu\pi\over
2}-{\pi\over 4} \right)} \,.
\end{equation}
The full intensity distribution  at $q_0$ is given by an energy average.  Since
the level density is constant for a map (and constant within the energy window
of interest for a continuous-time Hamiltonian system), and the random factors
$|r_\alpha|^2$ are uncorrelated with the smooth envelope $S(E)$, each moment of
the intensity distribution may be obtained as a product of a factor depending
on properties of the orbit and a universal factor associated with Gaussian
random fluctuations:
\begin{equation}
\overline{I^s} =\left ({1 \over 2\pi\hbar/T} \int_{E_1}^{E_1+2\pi\hbar/T} dE \;
\left ( S(E)\right)^s \right )\; \cdot \;\left (\overline {|r_\alpha|^{2s}}
\right ) \,.
\end{equation}
Note that the smooth envelope is periodic in $T$, so we need only to average
over the energy interval $E_1 \le E \le E_1+2\pi\hbar/T$.
In particular, using $\overline{I^2}={\cal I}(q_0)$, $\overline
{|r_\alpha|^4}=F$, and Eq.~(\ref{se}), we recover the result of
Eq.~(\ref{scaripr}) for the inverse participation ratio.

In the absence of time-reversal symmetry, $r_\alpha$ is a complex Gaussian and
the random factor in Eq.~(\ref{infact}) is exponentially distributed,
\begin{equation}
{\cal P}(|r_\alpha|^2)=e^{-|r_\alpha|^2} \,,
\end{equation}
while in the presence of time-reversal symmetry $r_\alpha$ is real
and its square follows a Porter-Thomas distribution,
\begin{equation}
{\cal P}_{\rm TRS}(|r_\alpha|^2)={1 \over \sqrt{2 \pi |r_\alpha|^2}}
e^{-|r_\alpha|^2/2}
\,.
\end{equation}
For definiteness, we focus on the generic situation of no time-reversal
symmetry.  At a fixed energy $E$, the distribution of $I=S(E)|r_\alpha|^2$ is
${\cal P}(I)= {1 \over S(E)} e^{-I/S(E)}$, and the combined intensity
distribution over all energies is
\begin{equation}
\label{pscar}
{\cal P}(I)={1 \over 2\pi\hbar/T} \int_{E_1}^{E_1+2\pi\hbar/T} {dE \over
S(E)} e^{-I/S(E)} \,.
\end{equation}

Several limits are of particular interest.  If the scarring is relatively weak,
i.e.  $S(E)-1 \ll 1$ in Eq.~(\ref{se}), we may expand Eq.~(\ref{pscar}) to
second order in $S(E)-1$ to obtain
\begin{equation}
\label{ppert}
{\cal P}(I) =\left (1+ \epsilon -2 \epsilon I + {\epsilon \over 2} I^2
+O(\epsilon^2) \right ) e^{-I} \,,
\end{equation}
in complete analogy with perturbative results for disordered systems, in the
limit of weak disorder~\cite{mirlin}.  Here
\begin{equation}
\epsilon = {1 \over 2\pi\hbar/T} \int_{E_1}^{E_1+2\pi\hbar/T} dE
 \; (S(E))^2={{\cal I} (q_0) \over F} - 1=
{ {\cal I}(q_0)-2 \over 2} \,,
\end{equation}
and for $\epsilon \to 0$ we recover the Porter-Thomas distribution
predicted by random matrix theory: ${\cal P}_{\rm RMT}(I)= e^{-I}$.
In the perturbative regime, deviations of the intensity distribution
from the RMT prediction are proportional to the deviation of the IPR
from its RMT value of 2.  We note that the validity of
Eq.~(\ref{ppert}) requires $I \ll 1$ in addition to $I \epsilon\ll
1$, so perturbation theory does not apply in the tail of the
intensity distribution.

To obtain a simple expression for the probability of rare events,
i.e. intensities much higher than the average intensity $\overline I
=1$, we need to follow a different approach.  For large $I$, the
integral in Eq.~(\ref{pscar}) may be evaluated using the saddle
point (stationary phase) method, with the dominant contribution
coming from the energy at which the smooth envelope $S(E)$ is
peaked.  Physically, this means that very large wave function
intensities will almost always be at the ``scar energies" $E_0$
where interference from successive iterations of the orbit add fully
constructively, and never at the ``antiscar energies" where this
interference is destructive. We then obtain
\begin{equation}
{\cal P}(I) = {1 \over 2\pi\hbar/T }
\sqrt{2 \pi  \over  |S''(E_0)| I} e^{-I/S(E_0)}
\end{equation}
where $E_0$ is the energy at which the smooth envelope $S(E)$ of Eq.~(\ref{se})
is peaked, and $S(E_0)$ and $S''(E_0)$ are the value and the second derivative
of $S(E)$ at the peak.

For small $\lambda$, we may rewrite the sum in Eq.~(\ref{se}) as an integral,
which may be evaluated in terms of hypergeometric functions.  We then find that
the smooth local density of states is peaked at
\begin{equation}
E_0\approx {-S+\mu\pi\hbar/2+\beta\lambda\hbar \over T} \,,
\end{equation}
where the dimensionless constant $\beta \approx  0.2133$ is the solution of
$\int_0^\infty dx \;x \sin(\beta x-\pi/4)/\sqrt{\sinh{x}}=0$.  The height of
the peak is given in this limit by
\begin{equation}
S(E_0)\approx \sqrt{2 \over N |ab|} { \gamma \over \lambda}\,,
\end{equation}
while the second derivative is
\begin{equation}
S''(E_0)\approx -\sqrt{2 \over N |ab|} {T^2 \over \hbar^2}
{ \delta \over \lambda^3}\,,
\end{equation}
where $\gamma=\int_0^\infty dx \;\cos(\beta x-\pi/4)/\sqrt{\sinh{x}}\approx
3.059$ and $\delta =\int_0^\infty dx \;x^2 \cos(\beta
x-\pi/4)/\sqrt{\sinh{x}}\approx 16.14$ are dimensionless constants.  Combining
these results, the tail of the intensity distribution for small $\lambda$
finally takes the form
\begin{equation}
\label{ptail}
{\cal P}(I) \approx \sqrt{ \lambda^3
\over 2\pi \delta A I} e^{-\lambda I/\gamma A}\,,
\end{equation}
where $A=\sqrt{2/N|ab|}$ is a geometrical factor that becomes $O(1)$ near a
focal point of the periodic orbit.  Comparing with ${\cal P}_{\rm
RMT}(I)=e^{-I}$, we note that the tail of the intensity distribution remains
exponential in the presence of scarring, but the exponent now depends on the
instability exponent $\lambda$, leading to a greatly enhanced probability of
finding very high wave function intensities on weakly unstable orbits.  The
exponentially small probability of rare events in {\it ballistic} chaotic
systems, as given by Eq.~(\ref{ptail}), contrasts with the log-normal tail
predicted and observed for {\it diffusive} two-dimensional systems in the
metallic regime~\cite{mirlin}.

In the opposite limit $I \to 0$, Eq.~(\ref{pscar}) implies ${\cal P}(I) \to
\overline{S^{-1}} > 1/\overline{S} =1 = {\cal P}_{\rm RMT}(I)$; this enhanced
probability of very small intensities is known as the antiscar effect.

\section{Numerical Model and Results}
\label{secnumer}

For the purpose of testing the general results obtained in
Sections~\ref{secipr} and \ref{secdistr}, we consider a specific
family of discrete-time kicked maps defined classically on a
toroidal phase space $(q,p) \in [0,1) \times [0,1)$.  The time
evolution for one step $(q_n,p_n)  \to (q_{n+1},p_{n+1})$ is given
by
\begin{eqnarray}
\label{classevol}
q_{n+1/2}&=&q_n+T'_1(p_n) \;\; {\rm mod} \;\; 1 \nonumber \\
p_{n+1}& =& p_n -V'(q_{n+1/2}) \;\; {\rm mod} \;\; 1 \\
q_{n+1}&=&q_{n+1/2} +T'_2(p_{n+1})  \;\; {\rm mod} \;\; 1 \nonumber
\end{eqnarray}
with
\begin{eqnarray}
\label{tvdef} T_1(p)&=& {1 \over 2} (p-p_0)^2
+ D_1\left [ 2\sin{2\pi (p-p_0)}- \sin{4\pi(p-p_0)} \right ] \nonumber \\
V(q)&=& -{1 \over 2} K_q (q-q_0)^2
+D_2 K_q \left [ 2\sin{2\pi (q-q_0)}- \sin{4\pi(q-q_0)} \right ] \\
T_2(p)&=&  {1 \over 2} K_p (p-p_0)^2 + D_3 K_p \left [ 2\sin{2\pi
(p-p_0)}- \sin{4\pi(p-p_0)} \right ] \,. \nonumber
\end{eqnarray}
Physically, such a map may be thought of arising from a time-periodic
Hamiltonian where a particle experiences free evolution under the influence of
kinetic term $T_1(p)$ for the first half of a period, followed by a sudden kick
of strength $V(q)$ and then by additional free evolution with kinetic term
$T_2(p)$ for the second half of the period.  The corresponding quantum
evolution for one step is given by
\begin{equation}
\hat U = e^{-i T_2(\hat p)/\hbar}e^{-i V(\hat q)/\hbar}e^{-i T_1(\hat p)/\hbar}
\,,
\end{equation}
where $\hbar =1/2\pi N$, and $N$ is the dimension of the Hilbert space.  If we
were interested in the spectral behavior only, we could of course view the same
dynamics stroboscopically right before or right after each kick, effectively
combining the two free evolution sub-steps into one governed by kinetic term
$T_1+T_2$.  In that case, choosing $q_0=p_0=0$ and integer values for $K_q$ and
$K_p$ (so that both $V'(q)$ and $T'_2(p)$ are continuous functions on the
torus) would correspond to a perturbed cat map, a system well studied in the
literature~\cite{pertcat}.

In our case, we choose non-integer values of $K_q$ and $K_p$, as well as
generic values for $q_0$ and $p_0$, resulting in diffraction at $q=0$ and
$p=0$.  Nonzero $q_0$ and $p_0$ also cause a breaking of time-reversal
symmetry.  By construction, the above map is fully chaotic, for positive $K_q$
and $K_p+1$, and has a periodic orbit of period $T=1$ at $(q,p)=(q_0,p_0)$. The
instability exponent $\lambda$ and unstable manifold orientation parameter
$|ab|$ of this orbit are implicitly determined by the parameters $K_q$ and
$K_p$ via the relations
\begin{eqnarray}
\label{kqkp}
2\cosh \lambda &=& 2+K_q(1+K_p) \nonumber \\
2|ab| \sinh \lambda &=& 1+K_p(1+K_q) \,.
\end{eqnarray}
Any desirable values of $\lambda$ and $|ab|$ may be obtained by selecting
appropriate $K_q$ and $K_p$ in accordance with Eq.~(\ref{kqkp}).  The
parameters $D_1$, $D_2$, and $D_3$ have no effect on the unstable orbit at
$(q_0,p_0)$ or on its monodromy matrix (Eq.~(\ref{monodromy})), and within
certain bounds these additional parameters do not affect the fully chaotic
nature of the dynamics~\cite{pertcat}.  Such additional parameters may be used
to construct a large ensemble of distinct classical systems with identical
linearized behavior near the dominant periodic orbit.  In our numerical
investigations, however, we did not see significant effects in wave function
statistics due to variation of $D_1$, $D_2$, and $D_3$, and for simplicity the
data shown below was collected using $D_1=D_2=D_3=0$.  For a given pair of
parameters $K_q$, $K_p$ (equivalently, $\lambda$, $|ab|$), we do generate an
ensemble of similar systems by varying $q_0$ and $p_0$, as well as varying
boundary conditions in both the $q$ and $p$ directions. Specifically, the
boundary condition variation is accomplished by taking parameters $\epsilon_1$
and $\epsilon_2$ uniformly distributed between $0$ and $1$, and requiring
$\Psi(q+1)=\Psi(q)\, e^{-i 2\pi \epsilon_1}$ and $\tilde \Psi(p+1)=\tilde
\Psi(p) \, e^{i 2\pi \epsilon_2}$; this is equivalent to the choice
$q_j=(j+\epsilon_2)/N, \;\; 0\le j \le N-1$ for the position-space basis and
$p_\ell = (\ell+\epsilon_1)/N, \;\; 0\le \ell \le N-1$ for the momentum-space
basis.

\begin{figure}[ht]
\centerline{\includegraphics[width=3.5in,angle=270]{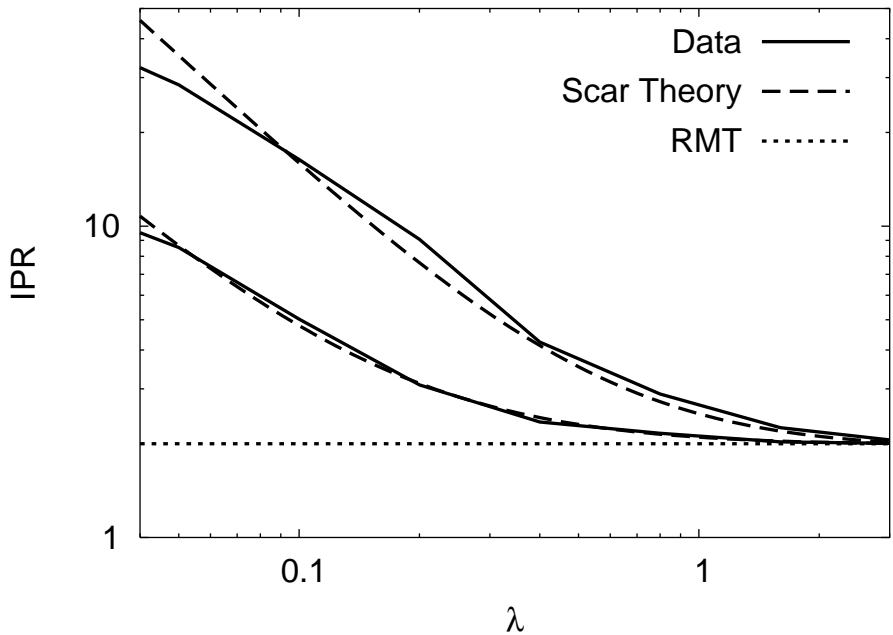}}
\vskip 0.2in
\caption{The average inverse participation ratio ${\cal I}(q_0)$ on a periodic
orbit is shown as a function of instability exponent $\lambda$ of the periodic
orbit, for the system defined by Eqs.~(\ref{classevol}) and (\ref{tvdef}) with
system size $N=1024$.  The upper and lower curves correspond to
$|ab|=0.005$, $0.025$, respectively.  The scar theory prediction of
Eq.~(\ref{smlamipr}) with $F=2$ is indicated by dashed lines.  All quantities
shown in this and subsequent figures are dimensionless.}
\label{fig_1}
\end{figure}

In Fig.~\ref{fig_1}, we examine the IPR as a function of the instability
exponent $\lambda$ of the periodic orbit, for two values of $|ab|$.  Good
agreement is observed with the simple asymptotic expression given by
Eq.~({\ref{smlamipr}) for a wide range of parameters, as the size of the
observed and predicted fluctuations varies by an order of magnitude.  We recall
that the validity of Eq.~(\ref{smlamipr}) requires $N^{-1} \ll \lambda \ll 1$, since
the orbit period $T=1$ in the present case.
Fig.~\ref{fig_2} shows what happens to the IPR as we leave the semiclassical
limit $N^{-1} \ll \lambda$.  We observe that to a very good approximation,
deviations from the semiclassical prediction are well described by a simple
scaling variable $\lambda N$:
\begin{equation}
\label{eqf}
{\cal I}(q_0)=f(\lambda N) \, {\cal I}_{\rm pred}(q_0) \,,
\end{equation}
where ${\cal I}_{\rm pred}(q_0)$ is the asymptotic prediction of
Eq.~(\ref{smlamipr}), $f(x) \to 1$ for $x \gg 1$, and $f(x) \sim x$ for $x \ll
1$ (as indicated by the dotted line in Fig.~\ref{fig_2}).  The dimensionless
parameter $\lambda N$ may equivalently be expressed as the ratio
$T_H/T_\lambda$, where $T_H$ is the Heisenberg time at which individual
eigenstates are resolved and $T_\lambda=1/\lambda$ is a Lyapunov time
associated with classical decay away from the periodic orbit.  It is noteworthy
that the crossover between the semiclassical behavior at large $\lambda N$ and
the saturated behavior at small $\lambda N$ occurs around $\lambda N \approx
40$ for all parameter values considered. This implies, for example, that in
applications to quantum dots in the Coulomb blockade regime, where $30 \le N
\le 70$ in some typical experiments~\cite{cbspacing}, asymptotic $N \to \infty$
expressions are inadequate for describing wave function intensity statistics in
``generic" chaotic mean-field potentials ($\lambda \sim 1$), and finite-$\hbar$
expressions must be used instead.  Accurate modeling of wave function
statistical properties in such systems is necessary for proper understanding
and interpretation of conductance peak spacing experiments~\cite{cbthy}.

\begin{figure}[ht]
\centerline{\includegraphics[width=3.5in,angle=270]{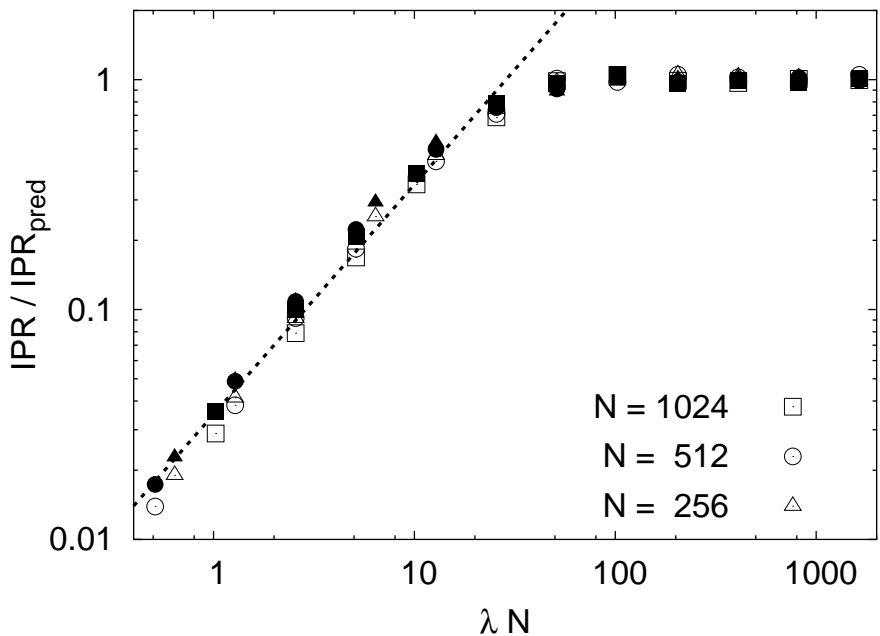}} \vskip
0.2in \caption{$f(\lambda N)$ (Eq.~(\ref{eqf})), the ratio of the
observed inverse participation ratio to the asymptotic prediction of
Eq.~(\ref{smlamipr}), is shown as a function of scaling parameter
$\lambda N$, for three different system sizes.  Open and filled
symbols represent $|ab|=0.025$ and $0.05$, respectively. The dotted
line indicates the asymptotic behavior $f(\lambda N) \approx 0.035
\,\lambda N$, for $\lambda N \ll 1$. } \label{fig_2}
\end{figure}

\begin{figure}[ht]
\centerline{\includegraphics[width=3.5in,angle=270]{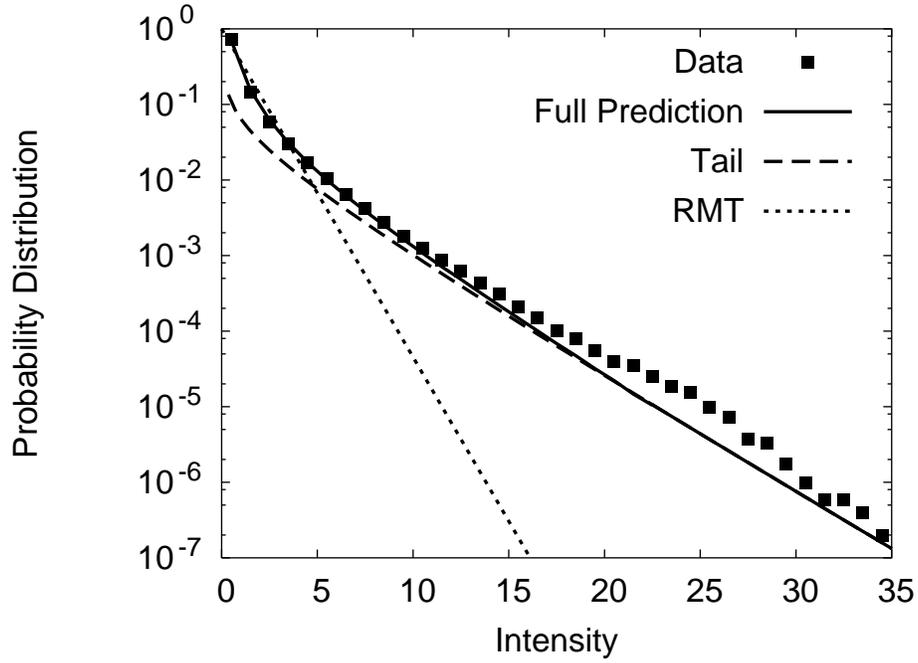}}
\vskip 0.2in
\caption{The wave function intensity distribution for $N=512$, $\lambda=0.9$,
and $|ab|=0.005$ is compared with the scar theory prediction of
Eq.~(\ref{pscar}) (solid line) and with the simpler expression of
Eq.~(\ref{ptail}) (dashed line), which is valid for large intensities. The
random matrix result ${\cal P}_{\rm RMT}(I) =e^{-I}$ is shown by a dotted line
for comparison.}
\label{fig_3}
\end{figure}

\begin{figure}[!h]
\centerline{\includegraphics[width=3.5in,angle=270]{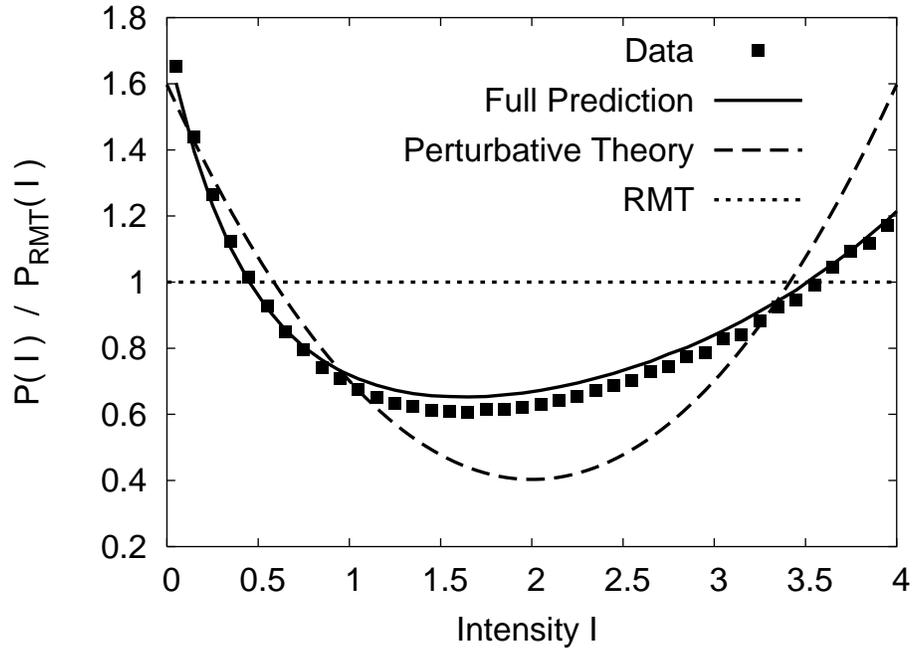}}
\vskip 0.2in
\caption{The wave function intensity distribution ${\cal P}(I)$ is
scaled by the random matrix theory prediction ${\cal P}_{\rm RMT}(I)$,
and the result is compared with the full prediction of Eq.~(\ref{pscar})
and the perturbative limit of Eq.~(\ref{ppert}).  All parameters are the same
as in the previous figure.
}
\label{fig_4}
\end{figure}

We now turn to the full distribution of wave function intensities. A
typical probability distribution is shown in Fig.~\ref{fig_3}, and
the scar theory prediction of Eq.~(\ref{pscar}) clearly describes
the data much better than does the random matrix theory result.  We
note that the distribution of rare high-intensity events is in
adequate quantitative agreement with the simple ``tail" formula of
Eq.~(\ref{ptail}), although the latter result is strictly valid only
in the $\lambda \to 0$ limit.  In Fig.~\ref{fig_4}, we see that the
scar theory prediction also works well in the region of small to
moderate intensities; in particular it correctly predicts the
antiscar enhancement for $I \ll 1$.  The parameters used here are
evidently outside the range of validity of the perturbative
expression (Eq.~(\ref{ppert})).  For the values of $N$ and $|ab|$
considered in Fig.~\ref{fig_4}, the perturbative result becomes
applicable in the main body of the distribution for $\lambda \ge
1.5$ (not shown), where the system is sufficiently unstable that
deviations from RMT fall below the $20\%$ level.

\section{Summary}
\label{secsum}

We have studied the distribution of position-space wave function intensities on
an unstable periodic orbit in a ballistic chaotic system. In the
two-dimensional case (equivalently, in a one-dimensional discrete-time map),
the full distribution of intensities is given by three parameters: (i) the
system size in wavelength units, or alternatively the dimensionless
conductance, inverse effective $\hbar$, or Heisenberg time in units of a
one-bounce time; (ii) the instability exponent of the periodic
orbit; and (iii) orientation of the unstable manifold or alternatively the
proximity to a focal point.  Generalization to higher dimensions is
straightforward.

The tail of the probability distribution is exponential, with exponent
proportional to a simple function of the above three parameters.  This
contrasts with the log-normal distribution of rare events in disordered
two-dimensional systems.  When the scar effect is weak, the behavior for
moderate intensities is given by a perturbative expression in complete analogy
with the disordered case.

Simple semiclassical expressions begin to break down when the classical
Lyapunov time associated with decay away from the periodic orbit reaches $2.5 -
3 \%$ of the Heisenberg time. A better quantitative understanding of these
saturation effects is needed for reliable predictions of wave function
statistics in finite-size ballistic systems.

\begin{acknowledgments}
This work was supported in part by the Louisiana Board of Regents Support Fund
Contract LEQSF(2004-07)-RD-A-29.
\end{acknowledgments}

\end{document}